\documentclass[trackchanges,twocolumn,twocolappendix]{aastex701}
\usepackage{enumitem}
\usepackage{amsmath}

\begin{document}

\title{The Hidden Evolution of Gamma-Ray Burst Jets Revealed by Gravitational Waves}

\author[orcid=0000-0002-7834-3113,sname='Urrutia']{Gerardo Urrutia}
\affiliation{Department of Astronomy and Astrophysics, University of California, Santa Cruz, CA 95064, USA}
\email[show]{geurruti@ucsc.edu}

\author[orcid=0000-0002-3137-4633,sname='De Colle']{Fabio De Colle}
\affiliation{Instituto de Ciencias Nucleares, Universidad Nacional Aut{\'o}noma de M{\'e}xico, A. P. 70-543 04510 D.F. Mexico}
\email{fabio@nucleares.unam.mx}

\author[orcid=0000-0002-1622-3036,sname='Janiuk']{Agnieszka Janiuk}
\affiliation{Center for Theoretical Physics, Polish Academy of Sciences, Al. Lotnikow 32/46, 02-668 Warsaw, Poland}
\email{agnes@cft.edu}

\author[orcid=0000-0003-2558-3102,sname='Moreno']{Claudia Moreno}
\affiliation{Departamento de F\'isica, CUCEI, Universidad de Guadalajara, 44430, Guadalajara, Jalisco, M\'exico}
\email{claudia@mail.com}  

\author[orcid=0000-0003-2558-3102,sname='Ramirez-Ruiz']{Enrico Ramirez-Ruiz}
\affiliation{Department of Astronomy and Astrophysics, University of California, Santa Cruz, CA 95064, USA}
\email{enrico@ukolick.org}  

\begin{abstract}
Gamma-ray burst (GRB) jets can propagate through massive stars, where the high opacity of the stellar envelope traps electromagnetic radiation and limits observations of the jet at early times. Gravitational waves (GWs), in contrast, can escape these dense environments and provide direct information of the jet dynamics. Previous studies have shown that GRB jets can emit GWs at low frequencies, while high-frequency emission may arise from jet-driven cocoons. However, the cocoon contribution typically requires energies much larger than those inferred from observations and produces stochastic fluctuations that can resemble noise. In this work, we show that high-frequency GW signals, can instead be generated by variability in the central engine. We consider jet models with different luminosity histories. We find that the GW emission depends on both the central engine history and the interaction of the jet with the stellar envelope. Rapid luminosity fluctuations are progressively suppressed as the jet propagates through the star, while the GW signal preserves information about the early jet evolution. Jets powered by strongly variable engines produce a GW signal peaking at tens of Hz, whereas smoothly varying engines are dominated by low-frequency emission. For a source at 36 Mpc, comparable to the distance of the closest GRB observed to date, the high-frequency signal from our most energetic variable models can reach amplitudes detectable by current ground-based interferometers. Our results suggest that the detection or non-detection of GWs from nearby GRBs and Type Ic broad-lined SNe could help constrain the mechanisms operating in the central engine.
\end{abstract}


\section{Introduction}
Gamma-Ray Bursts (GRBs) are thought to be produced by the dissipation of energy in relativistic jets resulting from binary neutron star mergers, black hole-neutron star mergers, or ``long GRBs'' during the collapse of rotating massive stars \citep[e.g.,][]{Gehrels2009,kumar15}. While the late-time dynamics of the jet is reasonably well understood, the mechanisms regulating the central engine and the initial stages of jet propagation are less clear. It is not possible to observe directly the central engine, due to the high opacity of its environment \citep[e.g.,][]{Woosley1993, MacFadyenWoosley1999, ramirez-ruizCocoon2002}. The properties of the central engine must be inferred from observations of the jet at late times, often with significant degeneracies in the physical interpretation \citep[e.g.,][]{Fabiodegeneracy2024}.

Gravitational waves (GWs) provide information on regions not observable by electromagnetic radiation. This was the case with GRB 170817A, the first example of a GRB observed together with a GW signal \citep{abbot2017NSmerger,coulter2017,kasen2017}. The GW emission constrains the neutron star masses, the final mass of the post-merger black hole, and its spin. Seconds later ($t\sim 1.74$~s), the prompt emission from the GRB was observed \citep[e.g.,][]{Murguia-Berthier2017,Gill_2019Collapse}, and its multi-wavelength afterglow emission was followed for months to years \citep{Margutti2018,Troja2018,MooleyNature2018,Makhathini2020}.

Relativistic sources can also produce GWs, although they produce a GW signal much weaker than those from compact binary mergers.
Solutions to the linearized Einstein field equations were presented by \citet{BraginskiiThorne1987} and \citet{Segalis_2001}. 
The dependence of the GW signal on the energy and acceleration of relativistic jets has been discussed in analytical works by \citet{sago04, Akiba2013,piran13, Piran2019,piran21,Soker2023Gws}. 
Structured jets has been studied by \citet{piran21}, while \citet{sago04,Huang2023} studied the deceleration stage. 
In these analytical models, the jet is described as a single particle or as a set of particles moving at relativistic speeds.

Numerical studies have connected the hydrodynamical evolution of relativistic jets to their GW emission.  
\citet{urrutia22GW} presented long-term numerical simulations of the GW signal produced by GRB jets propagating through the progenitor star and into the circumstellar material. They showed that the GW signal depends on the central engine lifetime and energy, the progenitor size, and the jet acceleration. \citet{urrutia22GW} estimated the detectability of the GW signals at low frequencies, showing that they will likely be detected by future space-based interferometers (e.g., DECIGO, BBO). 
In a subsequent study, \citet{gottlieb2022GW} showed that the jet cocoon can generate GW variability extending to relatively high frequencies, due to turbulence and asymmetric structures in the cocoon. These results show that GRB jets can produce GW emission at high frequencies, although the required cocoon energies may be larger than those typically observed in  GRBs.

Recent general relativistic magnetohydrodynamic (GRMHD) simulations of highly magnetized collapsars \citep[e.g.,][]{gottlieb2022a,gottlieb2022b,Urrutia2025Collapsar} find highly variable jet luminosities and a wide range of luminosity functions for GRB jets \citep[e.g.,][]{janiuk_variability2021,gottlieb2022a,gottlieb2022b,Urrutia2025Collapsar}.
This variability is also observed in GRB light curves and represents one of the main observational
probes of the central engine \citep{rees1994,sari1997,Kobayashi1997,Daigne1998,r-r-f2000,rr2002}. Long-GRB light
curves show complex temporal structure over a broad range of timescales \citep[e.g.,][]{norris1996},
yet their power-density spectra (PDSs) follow the simple
statistical behavior $P(\nu)\propto\nu^{-5/3}$ with a sharp high-frequency break near
$\approx 1$ Hz \citep{Beloborodov2000}. The origin of these characteristic
timescales remains uncertain, and may be related to the 
central engine \citep[e.g.,][]{janiuk_variability2021} or to the evolution of the relativistic outflow \citep{Beloborodov2000}. 

In the collapsar scenario, the jet must propagate across the stellar envelope before producing the observable GRB \citep{lopezcamara2013,Aloy2018Range,gottlieb2022a,Urrutia2025Collapsar}.
During this propagation, the jet decelerates and mixes with the stellar material, transferring energy and momentum to the surrounding
cocoon \citep{ramirez-ruizCocoon2002, Matzner2003}. The variability seen in the prompt gamma-ray emission may instead be a processed version of the original engine activity \citep[e.g.,][]{Aloy2000,MacFadyen2001}.

In this work, we address two related questions. First, we discuss whether rapid variability generated by the central engine of a collapsar can produce a GW signal detectable by current and future detectors. Second, we study how this variability is imprinted on the GW signal and how much of the temporal information generated by the central engine is preserved when the jet emerges from the star.
To address these questions, we compute GW signals produced by slowly varying jets and by rapid varying jet, which luminosities are extracted from GRMHD simulations of collapsars.

This work is structured as follows. In Section~\ref{sec:methods}, we describe our methodology, including the jet models and the improvements we made in the GW calculation with respect to our previous study \citep{urrutia22GW}. In Section~\ref{sec:results}, we present the results, focusing on geometric effects and the temporal evolution of the GW strain. In Section~\ref{sec:discussion}, we discuss the physical interpretation of the results. We emphasize how the progenitor affects the GW signal, and we discuss the prospects for future detections. Finally, in Section~\ref{sec:conclusions}, we summarize our main results.

\section{Methods}\label{sec:methods}

\subsection{Numerical setup}\label{sec:numericalsetup}

We perform a set of special relativistic hydrodynamical simulations of long GRB jets using the \emph{Mezcal} code \citep{decolle12,decolle2012c,decolle2012b}, which employs a shock-capturing scheme and adaptive mesh refinement (AMR). The computational domain consists of a two-dimensional (2D), axisymmetric region in cylindrical coordinates $(r,z)$, extending along both axes up to $5.6\times 10^{11}$~cm. We use $40\times 40$ cells at the coarsest level and $N_l=13$ levels of refinement. At the maximum refinement level, the smallest cell corresponds to a spatial resolution of $\Delta r_{\mathrm min}\approx \Delta z_{\mathrm min}\approx 3.7\times 10^6$~cm. The AMR grid is refined based on the density gradient criterion $|\nabla \rho|/\rho \ge 0.4$, allowing increased refinement as the jet propagates across the computational domain. We maintain the same maximum resolution throughout the evolution of the jet (see \citealt{urrutia22_3D}), to properly compute the GW signal.

In our simulations, the gas follows an adiabatic equation of state with an index $\gamma_{\rm gas} = 4/3$. The integration time is $t_{\rm fin} = 20$ s, which is sufficient to follow the propagation of the jet inside the progenitor star and beyond, into the circumstellar medium.

\subsubsection{The initial conditions of the GRB progenitor and jets}

We consider two progenitor models \citep{WoosleyHeger_2006}: the Wolf-Rayet star 16TI, with a mass of $M_\star = 11.45\,M_\odot$ and a radius of $R_\star = 9 \times 10^{10}$~cm, and the progenitor 12TH, with a mass of $M_\star = 9.23\, M_\odot$ and a radius of $R_\star = 4.5 \times 10^{10}$~cm. We remap the spherically symmetric progenitor density profile $\rho_\star(r)$ onto the computational grid for $r \leq R_\star$. Outside the stellar region ($r \geq R_\star$), we impose a stratified wind given by $\rho_{\rm w} = \dot{M}_{\rm w} / (4 \pi r^2 v_{\rm w})$, where $\dot{M}_w = 10^{-5} \;M_\odot$~yr$^{-1}$ and $v_w = 10^3$~km~s$^{-1}$ \citep[e.g.,][]{rr2005, Vink2011}. To avoid abrupt density jumps between the progenitor surface and the wind, we set $\rho=$~max~$(\rho_\star,\rho_{\rm w})$.

We launch a pressure-dominated jet \citep[e.g.,][]{urrutia22_3D} from the inner boundary located at $r_{\mathrm j}=5\times 10^8$~cm, and within the angular region $0 \le \theta \le \theta_j$, where the jet opening angle $\theta_j = 0.1$~rad is defined with respect to the $z$-axis. The jet has an initial Lorentz factor $\Gamma_j = 10$ and velocity $v_j = (1 - \Gamma_j^{-2})^{1/2} c$, where $c$ is the speed of light. We assume an asymptotic Lorentz factor $\Gamma_\infty = 100$. In addition, we include the prescription to suppress the plug instability described in Appendix~A of \citet{Urrutia2021ShortGRBS}.

The total energy injected into the jet is given by
\begin{equation}
   E_{\rm j,tot} = \int_0^{t_j} L_j(t)\, dt \,,
   \label{eqn:max_energy}
\end{equation}
where $t_j$ is the jet injection time and $L_j(t)$ is the jet luminosity. 
In all simulations, we set $t_j \lesssim 10$~s, corresponding to the typical free-fall time of bound stellar material \citep{MacFadyenWoosley1999,Aloy2000,Zhang2025}.
We fix an isotropic energy $E_{\rm iso} = 10^{54}$~erg \citep[e.g.,][]{Atteia2017,Aloy2018Range,Angulo2024,Lloyd-Ronning2026b}.
The beaming corrected jet energy is given as $E_{\rm j, tot} = E_{\rm iso} (1 - \cos \theta_j)$.

To evaluate the impact of the progenitor structure on the GW signals, we fix a constant energy injection rate $L_j(t) = E_{\mathrm{j,tot}} / t_j$ for two models with different progenitors:
\begin{itemize}
    \item \textbf{Model S1:} Uses the progenitor star 12TH. 
    \item \textbf{Model S2:} Uses the progenitor star 16TI. 
\end{itemize}

To evaluate the effect of smooth changes in the jet luminosity on the GW signal, we use the progenitor 12TH and the following models:
\begin{itemize}   
\item \textbf{Model A1:} The jet luminosity is giv
\begin{equation}
    L(t) = L_0\begin{cases}
        1 & \text{if } t \leq t_0\, , \\
        (t/t_0)^{-5/3} & \text{if } t_0 < t < t_j\, , \\
        0 & \text{if } t > t_j\, ,
        \end{cases}
    \label{eqn:modelA1}
\end{equation}
where $L_0=2E_{\rm j, tot}/[5-3(t_j/t_0)^{-2/3}]$, such that the total jet energy is $E_{\rm j, tot}$ (see eq.~\eqref{eqn:max_energy}), and $t_0=1$~s.

\item \textbf{Model A2:} The luminosity is defined as 
\begin{equation}
    L(t) = L_0 \begin{cases}
        (t/t_0) & \text{if } t \leq t_0\, , \\
        (t/t_0)^{-2} & \text{if } t_0 < t < t_j\, , \\
        0 & \text{if } t > t_j\, ,
    \end{cases}
    \label{eqn:model_A2}
\end{equation}
where $L_0=6E_{\rm j, tot}/(5-2~(t_j/t_0)^{-3})$ is chose so that the total jet energy (see eq.~\eqref{eqn:max_energy}). The luminosity history represents the growth phase up to the magnetically arrested
disk (MAD) state (assumed to occur at $t_0 = t_{\rm MAD} = 1$~s), and is motivated by GRMHD simulations \citep{Gottlieb2023LongShortUnifiedPicture}. 
\end{itemize}

To evaluate the impact of a rapid energy injection variability on GW signals, we use the same progenitor 12TH and luminosity histories obtained from GRMHD simulations:
\begin{itemize}

    \item \textbf{Model M1:} This model uses a remapped luminosity history obtained from the GRMHD simulation of a jet launched from the black hole horizon, named as ``m1-$10^{-1}$B0'' in Table~1 from \citet{Urrutia2025Collapsar}. The luminosity exhibits a rapid rise within milliseconds due to the early onset of the MAD state. Immediately afterwards, the luminosity decays as $L(t) \propto t^{-0.49}$ and shows variability on timescales $\Delta t\sim 10^{-3}-10^{-2}$~s. The jet injection time is limited to $t_j = 6.7$~s, corresponding to the duration of the GRMHD simulation.
    
    \item \textbf{Model M2:} The luminosity function is also taken from the GRMHD simulation presented as ``m12TH'' in \citet{Urrutia2025Collapsar}. This model also reach quickly the MAD state and shows rapid variability together with a post-peak luminosity $L(t) \propto t^{-0.54}$. Here, we assume the injection time $t_j = 6$~s, corresponding to the integration time of the GRMHD simulation. 
\end{itemize}
To  avoid numerical errors from large discontinuities in the physical variables, for all models we impose a gradual decrease in the jet velocity, as
$v(t) = v_j(t_j - t) / t_j$ for $t \geq 0.9\,t_j$.

\subsection{Calculation of the gravitational wave signal from GRB jets}
\label{sec:gw}

The GW emission is computed directly from the numerical simulations. In the calculation, each fluid element (i.e., a computational cell) is treated as a relativistic particle, and the total emission from the jet is obtained by integrating the contributions from all cells (for more details, see Appendix \ref{appA}). As a result, each value of $h(t_i)$ is determined by adding the contribution from more than one million computational cells. In \citet{urrutia22GW}, the hydrodynamical data were post-processed using $N_{\rm out}=600$ outputs. The integration time was $t_{\rm fin}=300$~s. The evolution of the strain was computed with a temporal resolution of $\Delta t=t_{\rm fin}/N_{\rm out}=0.5$ s. This corresponds to a sampling frequency of $\nu \approx 2$~Hz, and a Nyquist frequency of $\nu_{\rm Nyquist}\simeq 1$~Hz. 

To resolve higher frequencies in the GW strain, we require higher temporal resolution. This can be achieved in two ways. The first is to increase the number of outputs $N_{\rm out}$ saved from the simulations, i.e. saving data at every simulation time step ($1/\Delta t_{\rm sim}\sim 10^{4}$~Hz). This approach may require storing petabytes of data (cf. \citealt{gottlieb2022GW}).

Here, we compute the GW signal at high frequencies directly during the simulation (on-the-fly), rather than by post-processing. In this case, the maximum temporal resolution is determined by the simulation time step, which is given by $\Delta t_{\rm sim} \gtrsim N_c \Delta x/c$, where $N_c=0.4$ is the Courant number, $c$ is the speed of light, and $\Delta x$ is the size of the smallest cell. With our grid parameters (Section~\ref{sec:numericalsetup}), we get a maximum frequency of $\sim 1/\Delta t_{\rm sim}\sim  10^{4}$~Hz.

\section{From central engine variability to the GW signal}
\label{sec:results}

In this section, we discuss the GW signal generated by relativistic GRB jets, and how it is modified by the interaction of the jet with the progenitor star. We consider six jet models, in which we vary the progenitor stars, the luminosity history $L_j(t)$ and the rapid variability of the central engine. The details of hydrodynamic evolution of the jets propagating through the progenitor star are discussed in Appendix~\ref{appB}.

\subsection{The origin of the GW signal}

Unlike electromagnetic radiation from a relativistic jet, which is strongly beamed in the direction of motion, the GW signal is suppressed along the direction of motion, and it is maximum at a finite observing angle (this is known as the ``anti-beaming effect'', see, e.g., \citealt{piran13,urrutia22GW}).

\begin{figure}
    \centering
\includegraphics[width=1.0\linewidth]{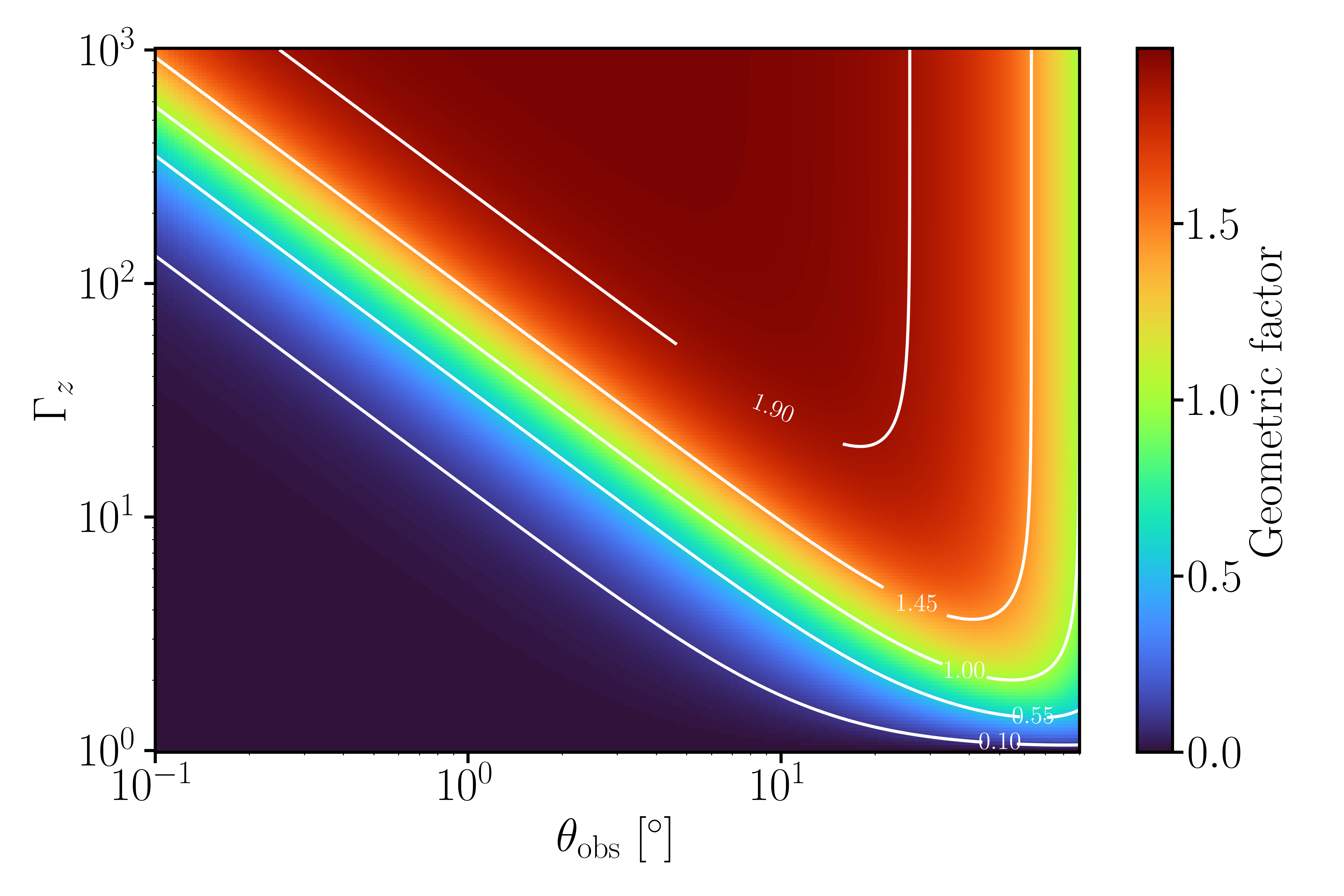}
    \caption{Angular dependence of the GW emission from a fluid element moving along the jet axis. We show the geometric factor $\beta_z^2\sin^2\theta_{\rm obs}/(1-\beta_z\cos\theta_{\rm obs})$, as a function of observing angle and jet Lorentz factor. The on-axis emission is suppressed for all Lorentz factors. As the Lorentz factor increases, the angle of maximum GW emission moves toward the jet axis.}
    \label{fig:geometricfactor}
\end{figure}

For the $h_+$ polarization, the angular dependence is regulated by the geometric factor
\begin{equation}
    \mathcal{G}_+ \equiv \frac{\beta^2\sin^2\theta_v}{1-\beta\cos\theta_v}\cos2\Phi\,.
    \label{eqn:geom_factor}
\end{equation}
For a source moving along the $z$-axis, $\cos\theta_v=\beta_z \cos\theta_{\rm obs}/\beta\lesssim 1$ and $\cos 2\Phi=1$ (see equations~\ref{eqn:costhv} and \ref{eqn:geom_fac}). Thus, for an on-axis observer, $\theta_{\rm obs}=0$, the geometric factor $\mathcal{G}_+ \rightarrow 0$, and the gravitational signal is $h_+ \sim h_\times \sim 0$.

The geometric factor is maximum at $\cos\theta_{v}=\cos\theta_{\rm obs}$ and $\cos2\Phi=1$. Taking the derivative of eq.~\eqref{eqn:geom_factor} with respect $\cos\theta_{\rm obs}$, we get an observing angle for which the signal peaks, i.e.,
\begin{equation}
    \cos\theta_{\rm obs}=\frac{\beta_z\,\Gamma_z}{\Gamma_z+1} = \sqrt{\frac{\Gamma_z-1}{\Gamma_z+1}}\, ,
\end{equation}
where $\beta_z=\sqrt{1-\Gamma_z^{-2}}$. At this observer angle, 
\begin{equation}
    \mathcal{G}_{+,{\rm max}}=\frac{\beta_z^2 \left(1-\cos^2 \theta_{\rm obs}\right)} {1 - \beta_z \cos \theta_{\rm obs} } 
    = 2\frac{\Gamma_z-1}{\Gamma_z}\,.
    \label{eqn:lor_obs_angle}
\end{equation}
In the limit $\Gamma_z\gg 1$, the signal is maximum for an observer angle $\theta_{\rm obs}\approx\sqrt{2/\Gamma_z}$, and the geometrical factor is $\mathcal{G}_{+,{\rm max}} \simeq 2$.

Figure~\ref{fig:geometricfactor} shows the dependence of the geometric factor with observing angle and Lorentz factor. For mildly relativistic fluid velocities, the maximum occurs at large observing angles and the geometric factor is small. For larger Lorentz factors, the maximum shifts towards smaller observing angles, although it never reaches $\theta_{\rm obs}=0^\circ$.

\begin{figure}
    \centering
    \includegraphics[width=1.0\linewidth]{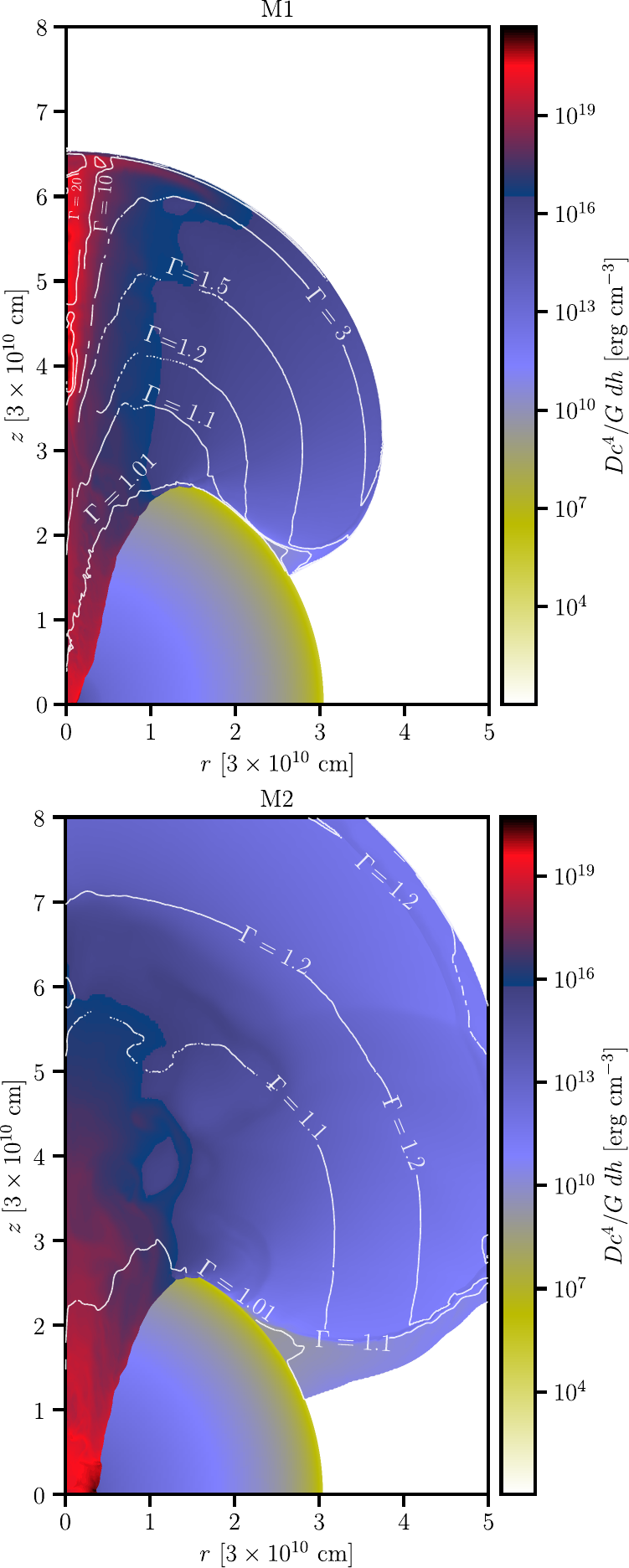}
    \caption{Two-dimensional maps of the GW emissivity at $t=10$ s, for models M1 (\emph{upper panel}) and M2 (\emph{bottom panel}). Model M2 produces a weaker and less relativistic jet than M1, concentrating a larger fraction of its GW emission in the inner energetic outflow.}
    \label{fig:jet_tracer_fig}
\end{figure}

To identify which part of the jet-cocoon system dominates the GW emission, we show in Figure~\ref{fig:jet_tracer_fig} two-dimensional maps of the GW signal at $t=10$~s, for models M1 and M2. For each computational cell, the map is computed at the observing angle that maximizes the geometric factor. In this case, the maximum strain contribution depends on the local energy $dE$ and Lorentz factor as
\begin{equation}
\frac{Dc^4}{G}~dh= \frac{4~(\Gamma-1)}{\Gamma}dE \, .
\label{eqn:max_h_per_volume}
\end{equation}
The map of Model M1 (upper panel) shows that the signal is dominated by fast-moving material located near the jet axis, consistent with \citealt{urrutia22GW}. The emission from the cocoon is $\sim$ five orders of magnitude lower, and the stellar material is not contributing significantly to the signal. Model M2 (lower panel) corresponds to a weak jet, i.e. a jet that reaches the stellar surface, but most of its energy is carried by mildly-relativistic velocities. In this model, most of the GW signal comes from the inner regions of the jet, where the high-energy material is concentrated.

\subsection{The imprint of central engine activity}

\begin{figure}
    \centering
\includegraphics[width=1.0\linewidth]{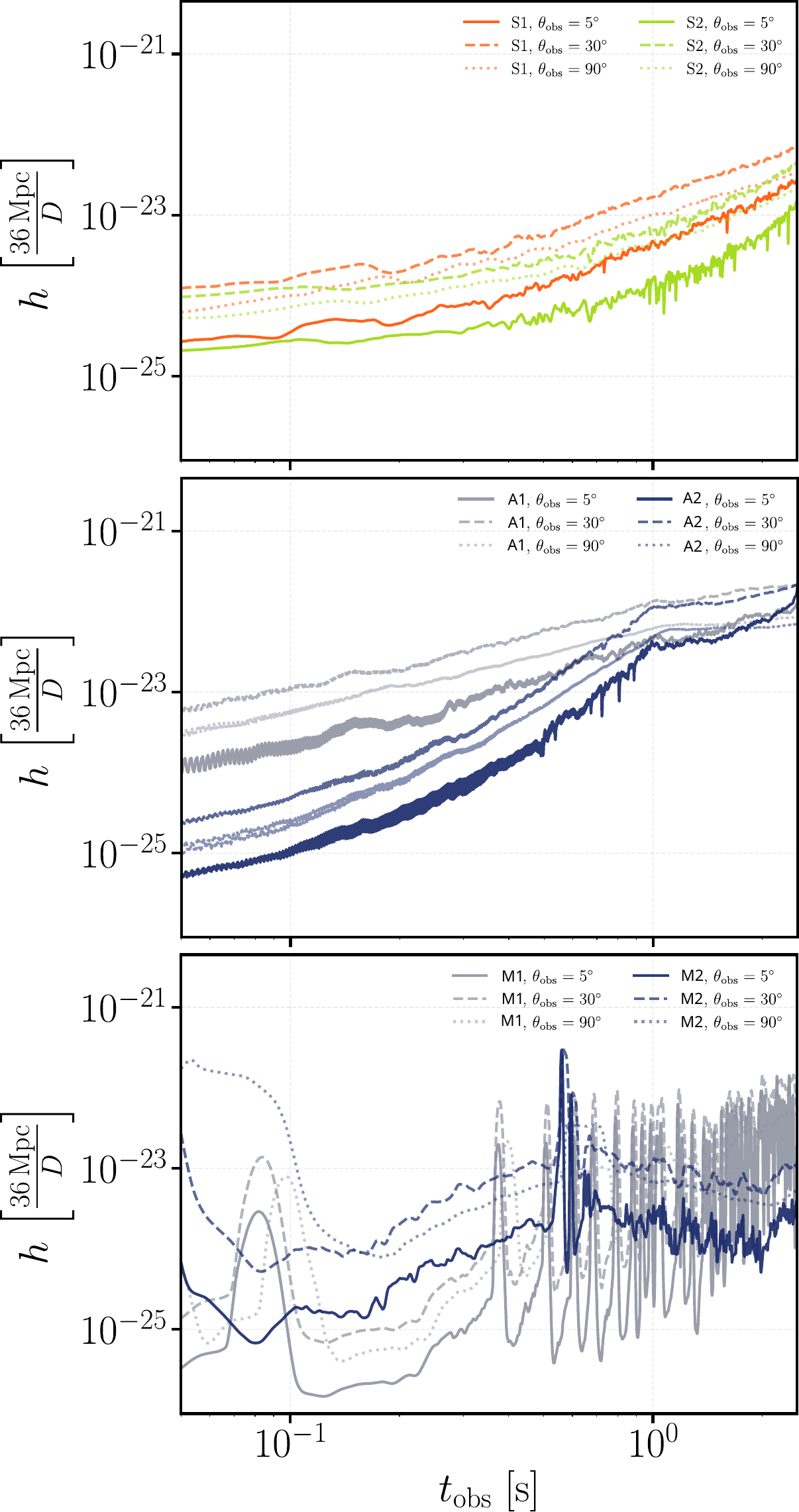}
    \caption{GW strain computed from the numerical simulations. \emph{Upper panel:} Effects of the progenitor structure on the strain evolution $h(t)$. Model S1 uses progenitor 12TH, while model S2 uses progenitor 16TI. The jet was injected with the same luminosity $L_j(t) = E_{\mathrm{tot}} / t_j$ and the same total energy in both models.
    \emph{Middle panel:} Effects of different energy injection profiles. Models A1 and A2 are injected with luminosities given by  eq.~\ref{eqn:modelA1} and ~\ref{eqn:model_A2} respectively.
    \emph{Bottom panel:} Strain evolution considering the energy injection profile imported from GRMHD simulations (models M1 and M2).}
    \label{fig:effectprogenitor}
\end{figure}

Figure~\ref{fig:effectprogenitor} presents the strain for the simulations introduced in Section~\ref{sec:methods}. Assuming a flat space-time, $H_0= 69.6$, $\Omega_M = 0.286$, and $\Omega_{\rm vac} = 0.714$,  we set a distance of $D=36$~Mpc, which corresponds to the nearest long GRB ever detected, the GRB~980425 observed at $z=0.0085$ \citep{Daigne2007}. The upper panel compares models S1 and S2, for which the central engine and total injected energy are identical but the stellar progenitors differ. The signal presents approximately the same amplitude at early times. However, starting from $t\sim 1$~s, the strain slope increases differently between these two models, demonstrating that the progenitor structure affects the evolution of the GW strain. Even though neither model contains variability in the injected luminosity, small-amplitude oscillations appear at $t\approx1$ s, when recollimation shocks develop in the jet funnel, and persist during the subsequent propagation. As the velocity evolution depends on the jet–cocoon pressure balance \citep[e.g.,][]{Hamidani2021-expanding,Urrutia2025Collapsar}, the progenitor modifies the strain, generating rapid variability in the GW signal even for a steady central engine, but with a modest amplitude compared with the GRMHD-driven models.

The middle panel shows the effect of the jet luminosity history $L(t)$ (see eq.~\ref{eqn:modelA1} and \ref{eqn:model_A2}) by comparing models A1 and A2 for the same 12TH progenitor. The behavior of $h(t)$ follows qualitatively the jet luminosity history. Model A1 rises rapidly while its luminosity remains high during the first second, while model A2 rises more gradually since the injected luminosity increases toward its maximum. At $t>1$~s, $h(t)$ increases much more slowly as the injected jet luminosity drops. The GW strain therefore retains information about how the jet is powered by the central engine even after the jet interacts with the stellar envelope.

Finally, the bottom panel shows $h(t)$ for relativistic GRB jets luminosities obtained from GRMHD simulations of LGRBs \citep{Urrutia2025Collapsar}. In models M1 and M2, rapid changes in the injected luminosity depend on how the material accretes and is reprocessed by the central engine, as well as on the inner structure of the progenitor \citep[e.g.,][]{gottlieb2022b,Urrutia2025Collapsar}. 

The resulting GW signals show much stronger fluctuations than the other models. The strain $h(t)$ exhibits oscillations of nearly one order of magnitude for model M1. The oscillations are clearly visible for observers at $\theta_{\rm obs}=5^\circ$ and $\theta_{\rm obs}=30^\circ$, but drop substantially for $\theta_{\rm obs}=90^\circ$. The rise of the GW signal $h(t)$ is similar to the one seen in model A2, in which the injected luminosity increases until it reaches an approximately steady state. In the case of model M2, $h(t)$ shows an early peak associated with an initial outburst from the central engine. The signal reaches a peak near $t \sim 1$~s and then decreases as the jet velocity drops due to its interaction with the progenitor star. 

\subsection{Changes in variability during jet propagation}\label{sec:frequencies}

In this section, we discuss how the variability originating from the central engine \citep[e.g.,][]{janiuk_variability2021} changes as the jet propagates through the star. As a reference, we select the model ``m1-$10^{-1}$B0'' from \citet{urrutia22_3D}, labeled in this work as M1. This jet was launched from the central-engine and followed until breakout. For the parameters adopted in these simulations ($M_{\rm BH}=5M_\odot$, $a_{\rm BH}=0.9$, and $T_{90}=6.7$~s), the minimum characteristic frequency is associated to the jet duration $\nu_{t_j}\simeq 0.172$~Hz, which also represents the minimum frequency expected in the GW signal \citep[e.g.,][]{Akiba2013,piran13,urrutia22GW}, while the maximum timescale associated with the rapid variability in $L(t)$ is $\nu_{\rm max}\simeq 1456$~Hz.

\begin{figure}
    \centering
    \includegraphics[width=1.0\linewidth]{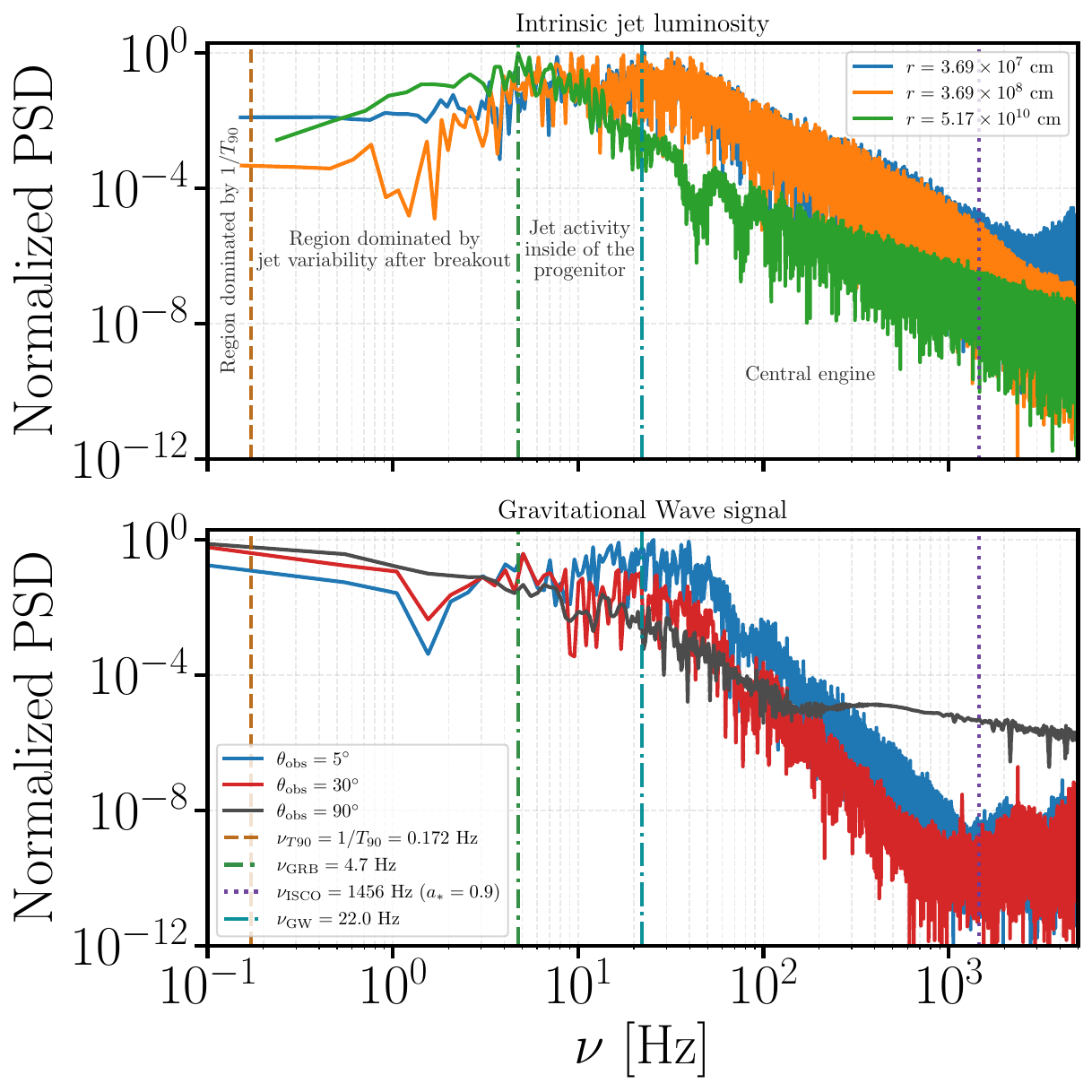}
    \caption{Evolution of the Power spectral density (PSD) of the jet luminosity and GW signal.
\emph{Upper panel}: PSD of the jet luminosity measured at three radii. Close to the central engine, the PSD extends over a broad range of frequencies. As the jet propagates outward, high-frequency variability is progressively suppressed and the spectrum becomes increasingly weighted toward lower frequencies. 
\emph{Lower panel}: PSD of the GW strain at three observing angles. The PSD extends over a broad frequency range.}
    \label{fig:psd_jet}
\end{figure}

Figure~\ref{fig:psd_jet} shows the power spectral density (PSD) of the jet luminosity at three radii (upper panel) and of the GW signal for thee viewing angles (bottom panel). Both the intrinsic jet luminosity and GW strain time series are normalized before computing their PSDs. The jet luminosity PSDs show the luminosity variability at different radii, while the GW PSD shows its dependence on the viewing angle. The luminosity $L_j(t)$ is computed at $3.69\times10^7$, $3.69\times10^8$, and $5.17\times10^{10}$~cm. The first two radii correspond to regions close to the central engine and within the dense stellar core, while the largest radius is close to the stellar surface. 

Close to the central engine, the luminosity contains significant power over a broad frequency range, extending from $\sim1$ to $500$~Hz, together with additional components at frequencies of order kHz. As the jet propagates outward, rapid fluctuations are suppressed, while variability on longer timescales ($1$-$50$~Hz) becomes increasingly important. By the time the jet approaches the stellar surface, its luminosity spectrum differs substantially from that near the jet base. This evolution results from the interaction of the jet with the stellar envelope, which modify the variability generated by the central engine. The emerging jet therefore retains only part of the variability injected at its base. This behavior is consistent with previous calculations showing that interactions with the progenitor can substantially modify jet variability before breakout \citep{lopezcamara2014,lopezcamara2016,Urrutia2025Collapsar}.

The PSD analysis reveals that the dominant luminosity variability frequency and the characteristic GW frequency are similar, with the PSDs peaking at $\nu_{\rm GRB}=19.8~{\rm Hz}$, and $\nu_{\rm GW}=22.0~{\rm Hz}$. This correspondence suggests that the strongest variability present in the jet can be preserved in the GW signal. The small difference between the two values may arise from dissipation in the jet. Finally, both $\nu_{\rm GRB}$ and $\nu_{\rm GW}$ are more than two orders of magnitude lower than $\nu_{\rm max}$. Therefore, the dominant spectral features of the jet luminosity and the GW signal are not directly associated with the central engine variability, but instead emerge from the subsequent evolution of the jet and its interaction with the progenitor star.

\begin{figure*}
    \centering
    \includegraphics[width=1.0\linewidth]{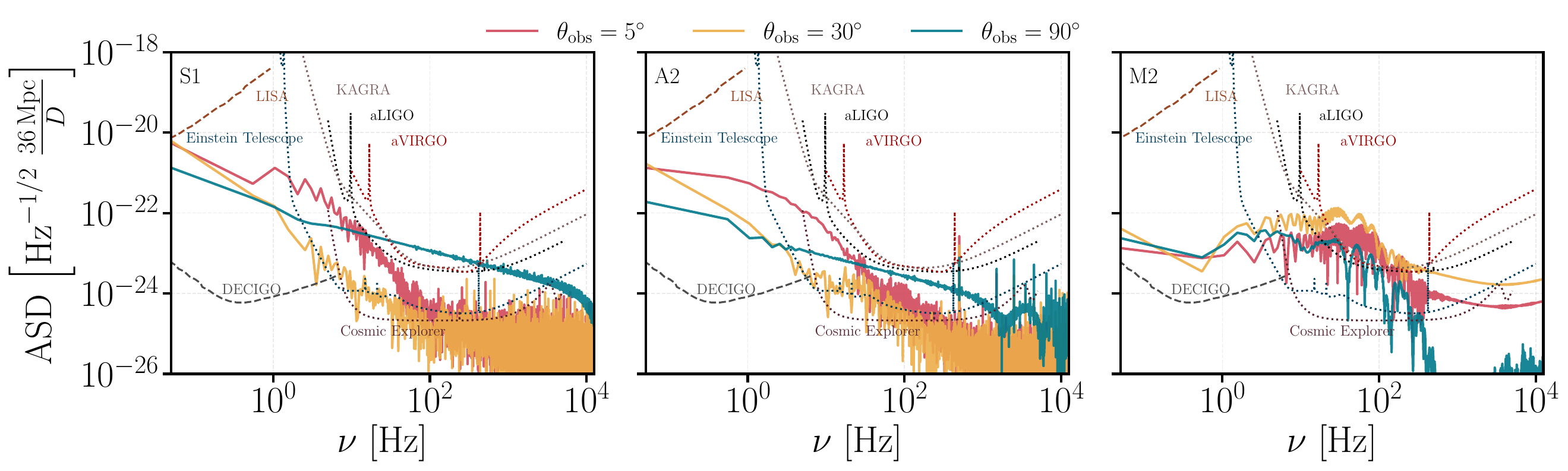}
    \caption{
    Amplitude spectral density (ASD) at a distance of $D=36$~Mpc for three representative jet models and three observing angles $\theta_{\rm obs}$, measured with respect to the jet axis. Model S1 has constant energy injection, A2 has a smoothly varying luminosity that peaks at $1$~s, and M2 uses a rapidly variable luminosity obtained from the GRMHD calculation. Sensitivity curves are shown for representative current and future ground-based interferometers (LIGO, Virgo, KAGRA, Einstein Telescope, and Cosmic Explorer) and space-based interferometers (LISA and DECIGO). In models S1 and A2, most of the GW power concentrates at low frequencies, whereas the variable M2 engine produces a broader spectrum with substantial power extending into the frequency range accessible to ground-based interferometers.
    }
    \label{fig:asd}
\end{figure*}

\section{Discussion}\label{sec:discussion}

In this work, we have analyzed how the strain evolution is shaped by the progenitor density structure, the jet luminosity history, and the variability of the jet. The total injected energy sets the overall scale of the signal, but the peaks and the temporal evolution of $L_j(t)$ determine the slope, oscillations, and spectral content of $h(t)$.

\subsection{Progenitor effects on the GW signal}

The progenitor structure mainly affects the rising phase of $h(t)$. Three elements are involved: the density profile of the progenitor, which determines the jet-head velocity \citep[e.g.,][]{Bromberg2011,harrison18,Hamidani2021-expanding}; the amount of energy transferred to the cocoon \citep[e.g.,][]{ramirez-ruizCocoon2002,decolle18b}; and the conversion of thermal energy into kinetic energy in the jet \citep[e.g.,][]{matsumoto2019,urrutia22_3D,Urrutia2025Collapsar}. 
Figure~\ref{fig:effectprogenitor} shows that, while the stellar density profile $\rho_\star(r)$ introduces a change in the normalization of the strain $h(t)$, the strain is strongly dependent on the jet luminosity history and on the presence of variability. 
Future observations of GWs associated with relativistic jets could then potentially provide direct constraints on the properties of the central engine which cannot be probed otherwise.

\subsection{Amplitude spectral density}

Using the strain evolution obtained from the numerical simulations, we estimate the amplitude spectral density 
\begin{eqnarray}
    {\rm ASD}=2\nu^{1/2}|\tilde{h}(\nu)|\; ,
\end{eqnarray} 
following \citet{Moore2014,urrutia22GW}. We keep a distance $D=36$~Mpc, which corresponds to the closest GRB~980425 \citep{Daigne2007} associated to a collapsing star, and three different observing angles: $\theta_{\rm obs}=5^\circ$, $30^\circ$, and $90^\circ$. 

Figure~\ref{fig:asd} shows the ASD for models S1, with constant energy injection, A2 with a smoothly evolving luminosity, and M2 whose luminosity profile is extracted from a GRMHD simulation and presents non-uniform variability, with several peaks associated with the onset of the magnetically arrested disk (MAD) state and the cessation of central engine activity. We also show in the figure sensitivity curves for present and future interferometers.

In models S1 ad A2, the ASD peaks at low frequencies, while in model M2, the ASD peaks at frequencies between $\nu\sim10^1$~Hz and $10^2$~Hz. At high frequencies, the ASD decreases and exhibits oscillations spanning approximately two orders of magnitude. These high-frequency variations may reflect physical variability that contributes only weakly to the observable GW signal, as high-frequency, stochastic variability associated with the central engine \citep{Sakai2025}.  

The amplitude of the signal also depends on the viewing angle. For models S1 and A2, the low-frequency component of the ASD is more prominent at low viewing angles, while the amplitude is larger for $\theta_{\rm obs}=90^\circ$. For model M2, the dependence on the observer angle is weak, as the outflowing material is less jetted.

\subsection{Detectability and observational implications}

\begin{figure}
    \centering
\includegraphics[width=1.0\linewidth]{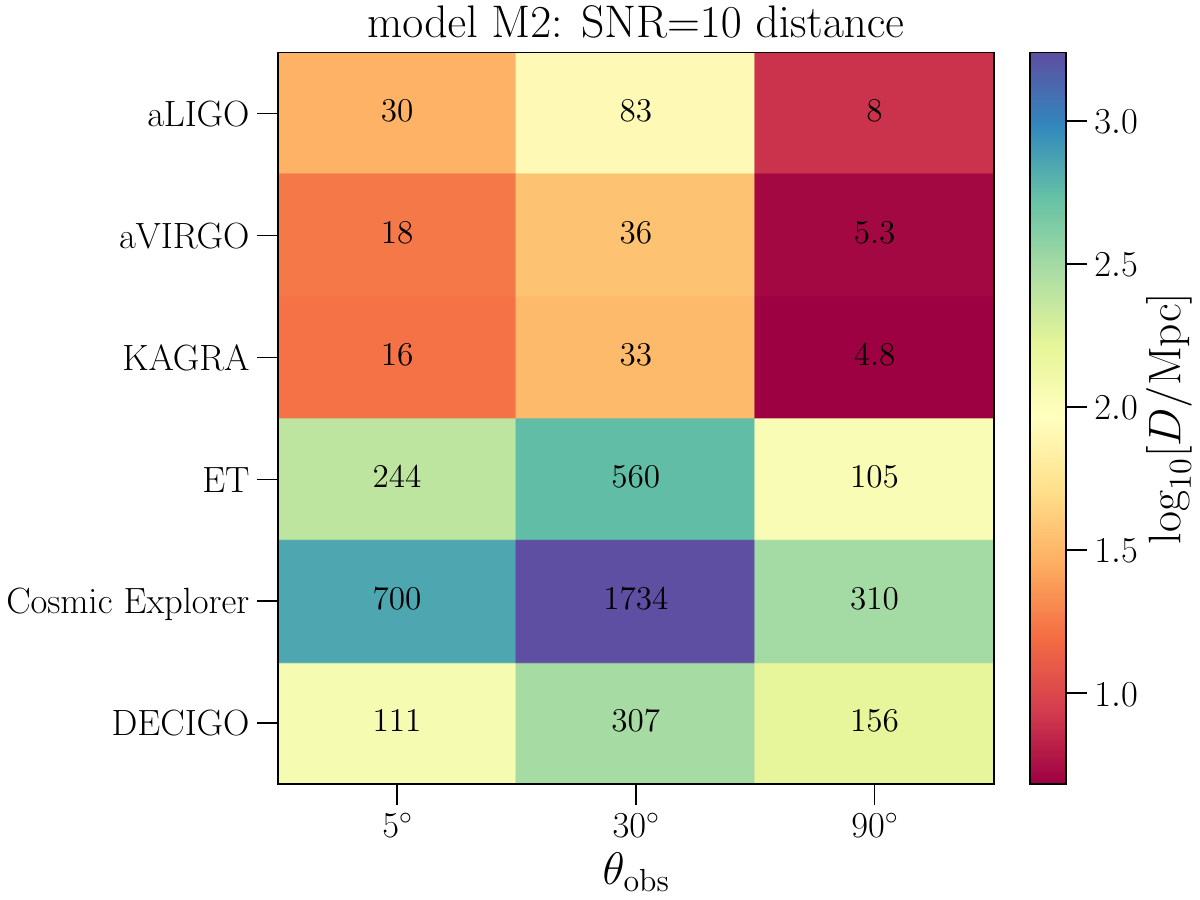}
    \caption{Maximum distance for detecting a GW signal from model M2 with ${\rm SNR}=10$ for three observing angles and different GW detectors. 
Model M2 is a weak jet obtained from GRMHD simulations, showing rapid variability. Current interferometers are restricted to nearby events, while next-generation detectors substantially increase the accessible distance. 
}
    \label{fig:detectable}
\end{figure}

Figure~\ref{fig:detectable} shows the maximum distance at which the GW signal from model M2 reaches a signal-to-noise ratio ${\rm SNR}=10$ for different GW observatories and observing angles. We focus on this model because its strong variability places a large fraction of the GW power within the frequency band of current ground-based interferometers, and also presents significant ASD at lower frequencies. The corresponding detection distance is of order of tens of Mpc for current ground-based detectors, several hundred Mpc for the Einstein Telescope \citep{einsteintelescope}, and up to Gpc scales for Cosmic Explorer \citep{cosmicexplorer} (with, in all cases, a strong dependence on the viewing angle). 

These results make weakly relativistic jets particularly interesting targets for GW searches \citep[e.g.,][]{Kaneko2007}. Nearby core-collapse supernovae may therefore contain a variable GW signal without an associated classical bright GRB. Current searches for GWs from nearby supernovae and GRB can thus provide constrains on central engine models, in a similar way to how GW searches are used nowadays to constrain supernova explosion mechanisms \citep[e.g.,][]{marek2024}. Nevertheless, this result should be treated with caution. If model M2 is considered a supernova-like candidate, its energetics lie toward the high end of the expected distribution.

\begin{figure}
    \centering
   \includegraphics[width=1.0\linewidth]{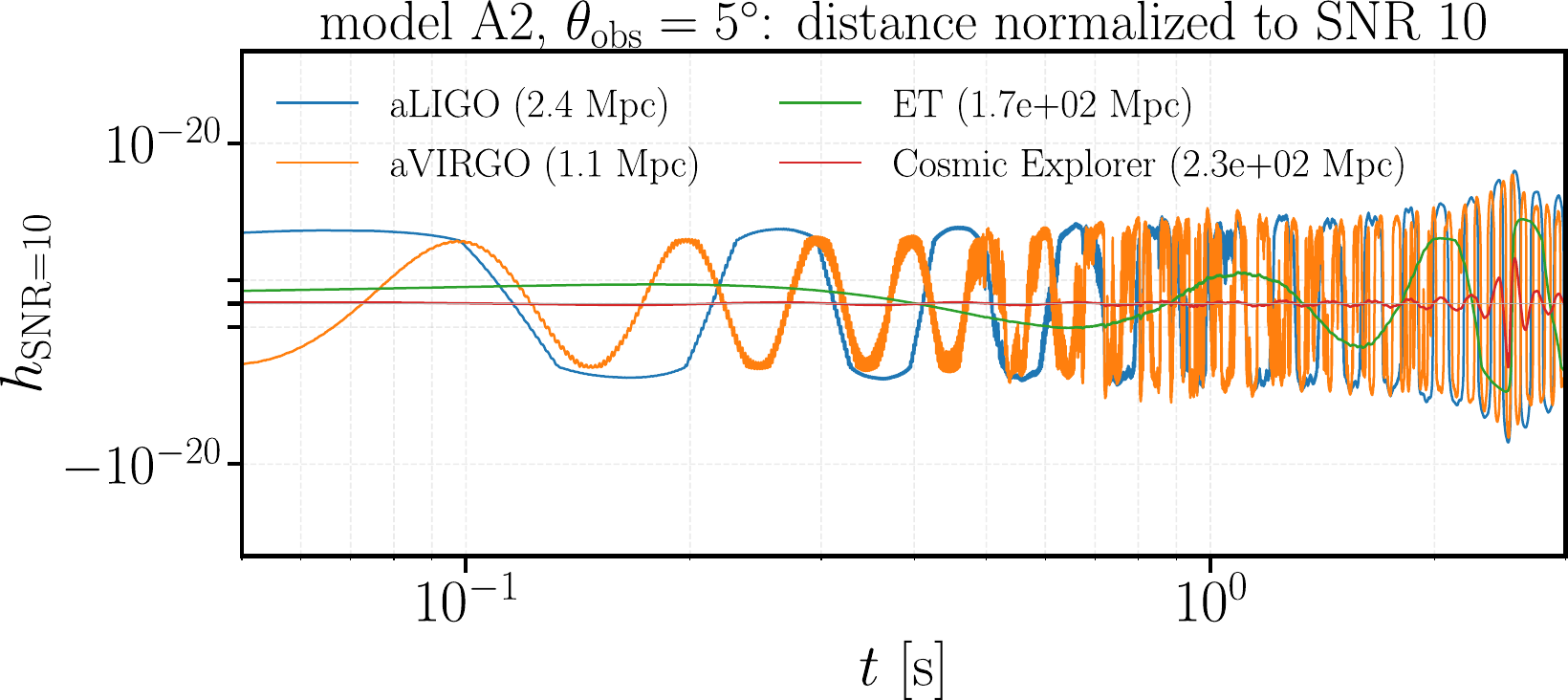}
    \includegraphics[width=1.0\linewidth]{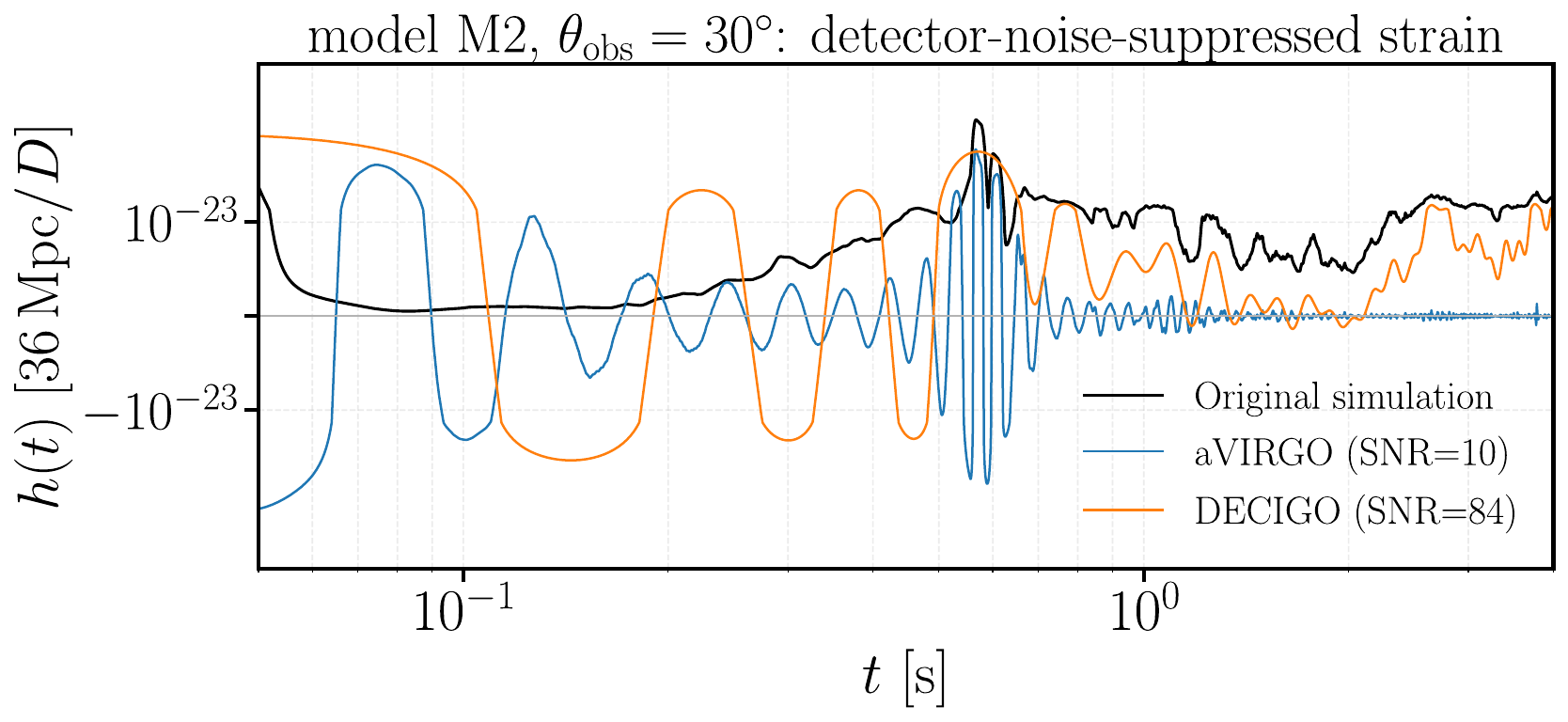}
    \caption{Strain signal $h(t)$ for models M2 (upper panel) and A2 (lower panel) after applying the representative detector sensitivity filtering used here at SNR = 10
from current and future ground-based interferometers. The figure illustrates how the simulated signals could be recovered. For model M2, the full temporal evolution of the strain is not recovered at high frequencies, however the peaks produced by the MAD activity remain visible above the noise level. Model A2 shows noise along its evolution.
    }
    \label{fig:noise_supress}
\end{figure}

As different interferometers are sensitive to different frequency ranges, only some features of the temporal evolution can be recovered by a single interferometer. In Figure~\ref{fig:noise_supress}, we show the strain signals computed from models A2 and M2 after accounting for the noise levels of current and future GW interferometers. That is, we neglect the signal below the noise level and consider an SNR of = 10. We define the signal-to-noise amplitude ratio between the simulation signal and detector noise as $r_k = {\rm ASD}_{\rm sim} (\nu)/{\rm ASD}_{\rm detector}(\nu)$, where the frequency range is set by the detector. We apply a Wiener filter $W = r^2 /(1+r^2)$, and obtain the detector-visible strain as the inverse Fourier transform $h_{\rm detector}(t) = \mathcal{F}^{-1} \{W\tilde{h}\}$.

The slowly varying component of the strain is concentrated at low frequencies and could be recovered by future detectors such as DECIGO \citep{decigopaper}. Model M2 instead produces high-frequencies components that may be detectable by ground-based observatories, although the full temporal evolution of the signal would not be recovered.

While the figure provides a more realistic representation of how the GW strain $h(t)$ could be observed, it should not yet be interpreted as a complete search strategy, since the reconstruction process employed here is based on a simplified signal-recovery calculation.

This study does not include other potential sources of GW emission in collapsars. Asymmetries in the accretion disk \citep{Gottlieb2024GWdisks,Fernandez2025GWdisks,Yuan2025GWdisks}, dynamics during the post-bounce phase and convection \citep[e.g.,][]{Vartayan2023Gws,Cusinato2026}, or  anisotropic neutrino emission \citep[e.g.,][]{Powell2024} may also produce GW signals. These mechanisms are expected to contribute at earlier times, whereas the jet emission studied here probes the subsequent central engine activity and jet propagation. GW emission associated with later activity \citep[e.g.,][]{Akiba2013,Sakai2025}, possibly associated with X-ray flares during the afterglow, has also been considered \citep{Huang2023}. However, the prospects for detecting such signals remain uncertain because their long timescales shift most of the GW power toward very low frequencies, where suitable detector sensitivity is limited.

\section{Conclusions}\label{sec:conclusions}

We have investigated the gravitational-wave emission produced by relativistic jets propagating through collapsar progenitors, with particular emphasis on the imprint of central engine injection. 
Using relativistic hydrodynamical simulations with an on-the-fly module to compute the GW strain, we have compared different jet models and analyzed the effects of the stellar progenitor and jet luminosity history on the GW.

Our main results are the following: 
\begin{enumerate}[noitemsep,nolistsep]

\item 
Jets with smoothly varying energy injection are dominated by low-frequency emission, potentially detectable by future low-frequency interferometers such as DECIGO, which could probe the global energy-injection history of the jet. In contrast, jets powered by MAD and highly variable central engines peak at $\nu \sim10^{1}-10^{2}$~Hz, within the sensitivity bands of ground-based interferometers, and could potentially be detected at distances of tens of Mpc with current detectors, several hundred Mpc with the Einstein Telescope, and up to Gpc scales with Cosmic Explorer. 

These estimates correspond to a highly energetic model, and the detection distances and event rates must be rescaled according to the jet-energy distribution of the source population under consideration. Nevertheless, the absence of GW detections associated with nearby GRBs and type Ic, broad-line SNe can be used to constrain the mechanism operating in central engines, similarly to what is routinely done for supernovae \citep[e.g.,][]{Marek2021,Gill_2022SNGW}.

\item 
Strong central engine variability produces a broader GW spectrum. The GRMHD-derived engines demonstrate the same behavior, with rapid variability associated with highly magnetized accretion and the MAD state. As the jet propagates, these fluctuations are progressively smoothed out in the jet luminosity, while the GW signal retains the signature of the central engine variability.

\end{enumerate}

The present study follows only the first $\sim20$~s of the jet-progenitor interaction and therefore does not fully characterize the very-low-frequency emission produced by longer-lived or ultra-long central engines. It also does not include other possible GW sources in collapsars, such as post-bounce dynamics, accretion-disk asymmetries, convection, or memory generated by anisotropic neutrino emission. Future simulations combining these channels over longer timescales will be required to construct a complete GW description of collapsar evolution.

\begin{acknowledgments}
We thank Pablo Cerdá-Durán, Michele Zanolin, Marek Szczepanczyk, Mikołaj Korzyński, Bozena Czerny, Piotr Plonka and Yuri Levin for useful discussions. This work was partially supported by ``Young Scientist Competition'' under the project titled ``Gravitational Wave Memory from GRB Jets'' sponsored by Center for Theoretical Physics (Polish Academy of Sciences), in Warsaw Poland. GU acknowledges support from University of California-Alianza MX Research \& Innovation: UCPRF2025-01. GU also acknowledges the support of the Heising-Simons Foundation and the Vera Rubin Presidential Chair at the University of California, Santa Cruz. GU and AJ acknowledge support by Polish National Science Center under the grant 2023/50/A/ST9/00527. GU and FDC acknowledge UNAM-PAPIIT grant IN113424. We gratefully acknowledge the computing time granted by DGTIC UNAM on the supercomputer Miztli (project LANCAD-UNAM-DGTIC-281) and the Polish high-performance computing infrastructure PLGrid (HPC Center: ACK Cyfronet AGH) for providing computer facilities and support within the computational grant PLG/2025/018086.  This research was also made possible through funding from the Lamat Institute \citep{2025NatAs...9.1770Q} and UC Santa Cruz through the Heising-Simons Foundation, and NSF grants AST-1852393, AST-1911206, AST-2150255, AST-2206243 and 2447606.
\end{acknowledgments}

\bibliographystyle{aasjournalv7}
\bibliography{main} 

\appendix

\section{Gravitational wave signal}
\label{appA}

The GW signal from a relativistic source can be derived from the linearized Einstein field equations \citep{BraginskiiThorne1987,Segalis_2001,Akiba2013}. We begin with a small perturbation $h_{\mu\nu} = g_{\mu\nu} - \eta_{\mu\nu}$ of the metric $g_{\mu\nu}$, where $\eta_{\mu\nu} = \text{diag}[-1,1,1,1]$ is the Minkowski metric. Under the Lorentz gauge condition $\bar{h}^{\mu\nu}_{,\nu} = 0$, the linearized Einstein field equations for a weak gravitational field are given (in geometrical units, i.e. $G=c=1$) by
\begin{equation}
  \left(-\frac{\partial^2}{\partial t^2} + \nabla^2 \right)\bar{h}_{\mu\nu} = -16\pi\, T_{\mu\nu}\;,
  \label{eqn:wave_eq}
\end{equation}
where $\bar{h}_{\mu\nu} = h_{\mu\nu} - \frac{1}{2} \eta_{\mu\nu} h^\lambda_{\ \lambda}$, and $T_{\mu\nu}$ is the energy-momentum tensor. For a point mass $m$ moving along a worldline $r^\alpha(\tau)$, where $\tau$ is the proper time and $r^\alpha(\tau)$ is the particle’s position in Cartesian coordinates, the energy-momentum tensor is given by
\begin{equation}
  T^{\alpha\beta} = \int m\, u^{\alpha}(\tau) u^{\beta}(\tau)\, \delta^{(4)}[x - r(\tau)]\, d\tau \;,  
\end{equation}
where $u^\alpha = dr^\alpha/d\tau$ is the particle four-velocity. The retarded solution of the wave equation \eqref{eqn:wave_eq} for the source term is given by the generalization of the Li\'enard-Wiechert potentials,
\begin{equation}
    \bar{h}^{\alpha\beta} = -4m \frac{u^{\alpha}(\tau)\,u^{\beta}(\tau)}{-u_\gamma(\tau)\,[x-r(\tau)]^{\gamma}}\Bigg|_{\tau=\tau_{\rm ret}} \,,
    \label{eqn:gen_sol}
\end{equation}
where $\tau_{\rm ret}$ is the retarded time, defined as the intersection between the world-line $r(\tau)$ and the observer’s past light cone. 

The metric perturbation in the Lorentz gauge is given by $h^{\alpha\beta} = \bar{h}^{\alpha\beta} - \frac{1}{2}\eta^{\alpha\beta}\bar{h}_{\gamma}^{\ \gamma}$. Substituting eq.~\eqref{eqn:gen_sol}, the metric perturbation becomes
\begin{equation}
    h^{\alpha\beta} = \frac{4m}{-u_\gamma(\tau)\,[x-r(\tau)]^{\gamma}} \left[ u^\alpha(\tau)u^\beta(\tau) + \frac{1}{2}\eta^{\alpha\beta} \right] \,.
    \label{eqn:gauge_metric}
\end{equation}
To obtain the GW signal, the metric perturbation \eqref{eqn:gauge_metric} is transformed from the Lorentz gauge to the transverse-traceless (TT) gauge. This gauge  includes only the spatial components, with $h_{t\mu}^{\rm TT} = 0$. 

In the TT gauge, the polarization components have been derived by \citet{Segalis_2001,Akiba2013,piran13} and are given by
\begin{eqnarray}
    h_+ \equiv h_{xx}^{\rm TT} = -h_{yy}^{\rm TT} &=& \frac{2G}{c^4} \frac{E}{D} \frac{\beta^2 \sin^2 \theta_v}{1 - \beta \cos \theta_v} \cos 2\Phi \;, 
    \label{eqn:new_h+}
    \\
    h_\times \equiv h_{xy}^{\rm TT} = h_{yx}^{\rm TT} &=& \frac{2G}{c^4} \frac{E}{D} \frac{\beta^2 \sin^2 \theta_v}{1 - \beta \cos \theta_v} \sin 2\Phi \;,
    \label{eqn:new_hx}
\end{eqnarray}
where $G$ is the gravitational constant, $E$ is the energy of the particle, $D$ is the distance between the source and the observer, and $\beta = v/c$ is the particle's velocity normalized to the speed of light. The polarization $h_+$ given by eq.~\ref{eqn:new_h+} dominates over $h_\times$ due to the axisymmetry of the problem  \citep{urrutia22GW}.
The simulations compute the energy density and velocity of each fluid element. The GW signal is then obtained by integrating the contribution from all cells.
This prescription has been used in analytical works to estimate the GW emission from the propagation of GRB jets \citep{sago04,Akiba2013,piran13}.

The direction of propagation of each fluid element is given by its velocity vector $\hat\beta$ in the lab frame, while $\hat{n}$ represents the observer's line of sight. The angle $\theta_v$ between these vectors is defined as 
\begin{equation}
      \cos\theta_v = \hat{n} \cdot \hat{\beta} = \frac{\beta_R \sin\theta_{\mathrm obs} \cos\phi + \beta_z \cos\theta_{\mathrm obs}}{\beta}\, .
      \label{eqn:costhv}
\end{equation}

The geometrical factor $\cos(2\Phi)$, which relates the observer frame to the lab frame, is given by
\citep[see,][]{Akiba2013,urrutia22GW}
\begin{equation}
\cos(2\Phi)=\frac{\left(\sin\theta\cos\phi\cos\theta_{\rm obs}-\cos\theta\sin\theta_{\rm obs}\right)^2-\sin^2\theta\sin^2\phi }{1-\left(\sin\theta\cos\phi\sin\theta_{\rm obs}-\cos\theta\cos\theta_{\rm obs}\right)^2}\,.
    \label{eqn:geom_fac}
\end{equation}
We define the time array $\{t_i\}_{\rm obs}$, defined as a function of the cell position
\begin{equation}
    t_{i,{\rm obs}} = t_i - (R/c) \cos\phi \sin \theta_{\mathrm obs} - (z/c) \cos \theta_{\mathrm obs} \;.
    \label{eq:tobs}
\end{equation}
We set the maximum time equal to the integration time of the simulation, i.e. $t_{\rm lab,max}=t_{\rm obs,max}=t_{\rm fin}=20$~s. The array has a size $N_t=5\times10^5$. Then, the time bins in the array are equally spaced, $\Delta t_{\rm lab}= \Delta t_{\rm obs} = 4\times 10^{-5}~$s, and $\nu_{\rm Nyquist}\simeq1.25\times10^{4}$~Hz. We also define arrays $\{h_+(t_i)\}_{\rm obs}$ and $\{h_+(t_i)\}_{\rm obs}$, sampled on the same time bins and updated at each simulation timestep. 

The GW signal is assigned by distributing the contribution of a cell over all observer time bins that overlap the emission interval. The signal from a cell is assumed constant over a step $[t_1,t_2]$ of duration $\Delta t=t_2-t_1$, and the $i$-th observer bin is centered at $t_i$ with width $\Delta t_{\rm obs}$, i.e., $\left[t_i-\Delta t_{\rm obs}/2,\,t_i+\Delta t_{\rm obs}/2\right]$, then the deposited fraction in bin $i$ is
\begin{equation}
f_i=\max\left(0, 
\frac{\min \left(t_2,t_i+\frac{\Delta t_{\rm obs}}{2}\right)-\max\!\left(t_1,t_i-\frac{\Delta t_{\rm obs}}{2}\right)}{\Delta t}\right).
\end{equation}
The contribution added to that bin is therefore the instantaneous source amplitude multiplied by $f_i$. For bins fully contained within $[t_1,t_2]$, this reduces to $f_i=\Delta t_{\rm obs}/\Delta t$, while the first and last overlapping bins receive partial weights corresponding to their actual overlap lengths. By using this method, the total contribution over each timestep is conserved, since $\sum_i f_i = 1$.

\begin{figure}
    \centering
    \includegraphics[width=1.0\linewidth]{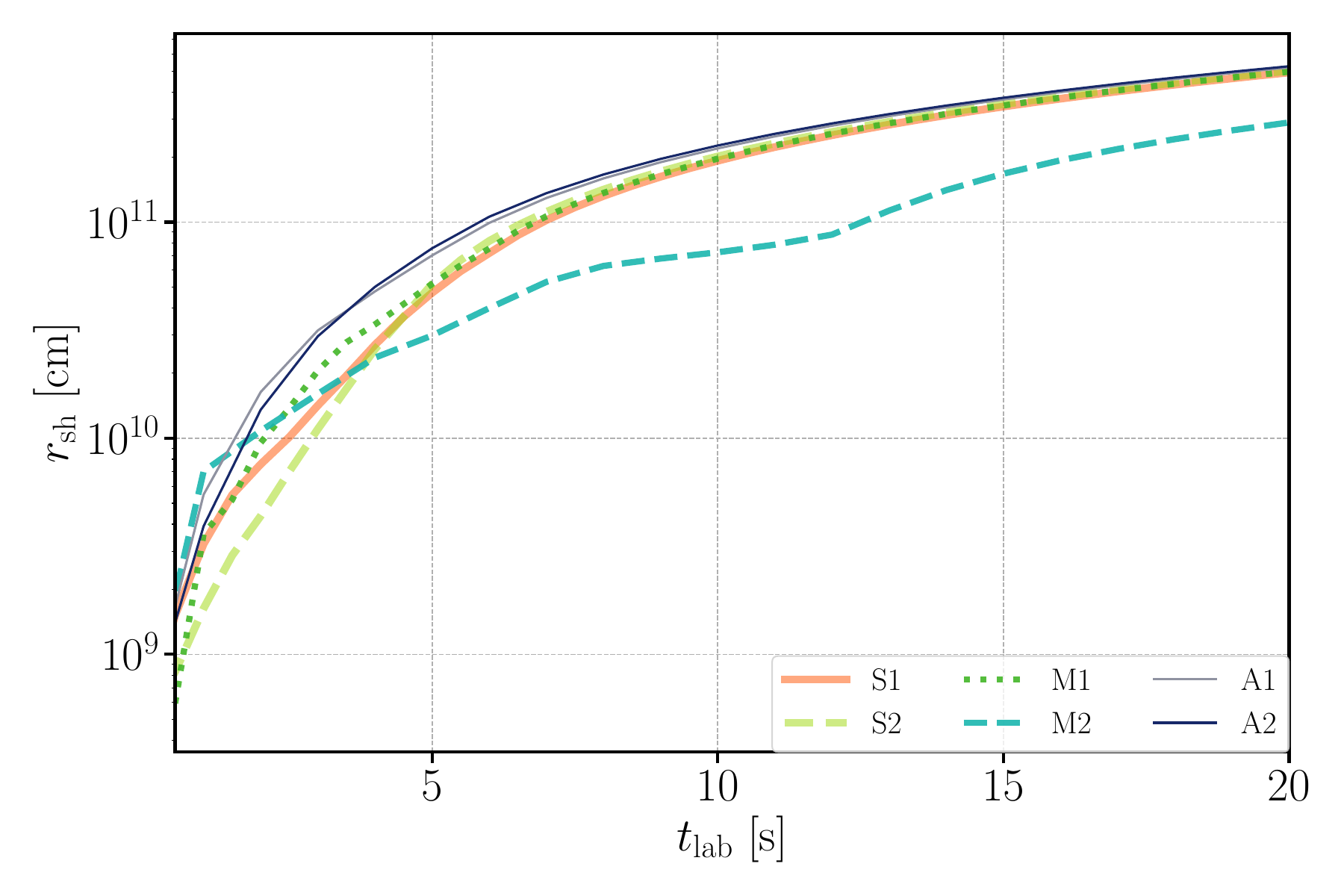}\\
    \includegraphics[width=1.0\linewidth]{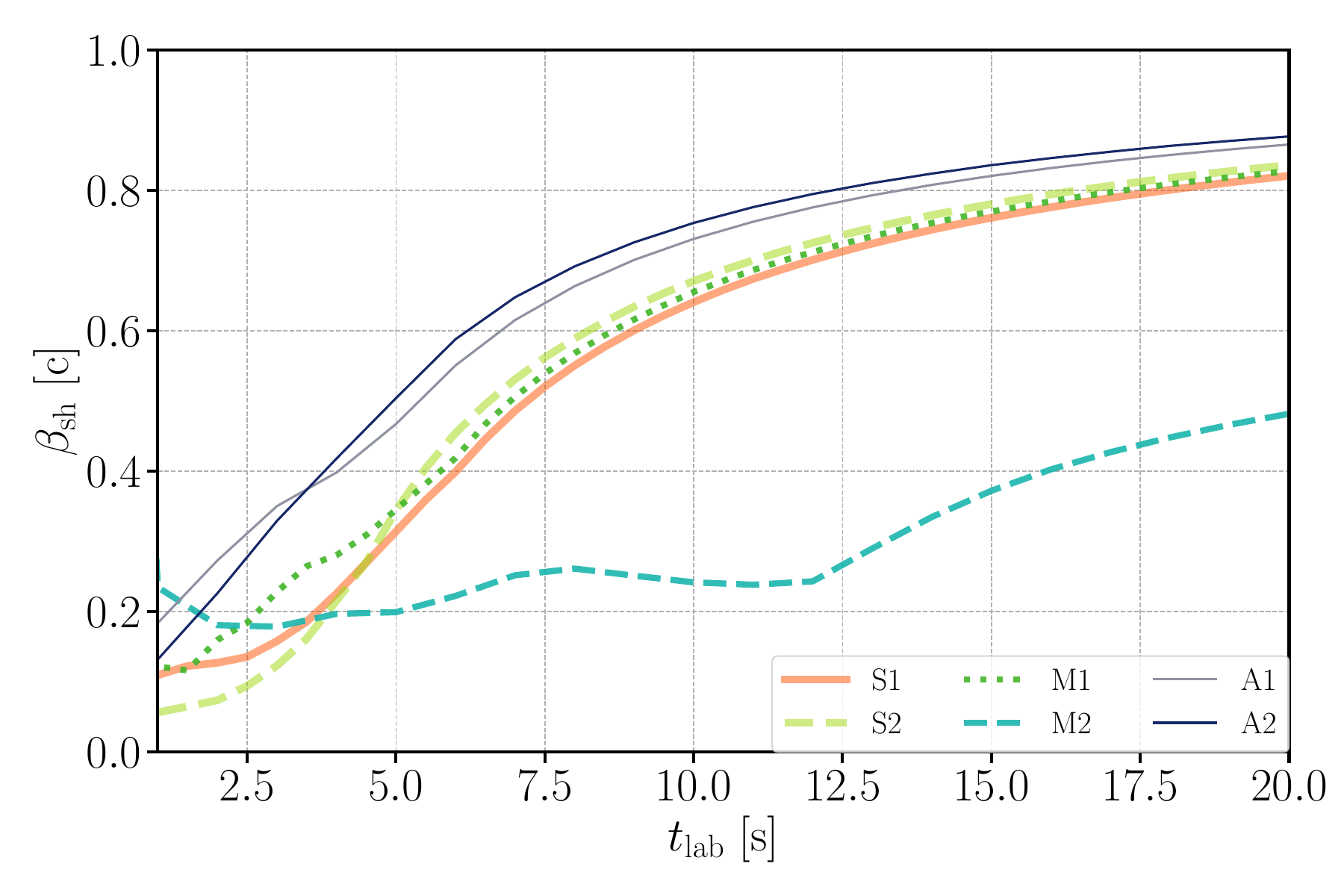}
    \caption{Position of the shock front (\emph{upper panel}) and jet head velocity (\emph{lower panel}) as a function of time, for the simulated models. Differences in progenitor structure and central engine history produce distinct propagation histories while the jets are inside the star. Most models subsequently accelerate after reaching the outer stellar layers. Model M2 remains substantially slower than the other models and develops only a weakly relativistic outflow, despite reaching the stellar surface.
}
    \label{fig:shock_beta}
\end{figure}

\begin{figure}
    \centering
    \includegraphics[width=1.0\linewidth]{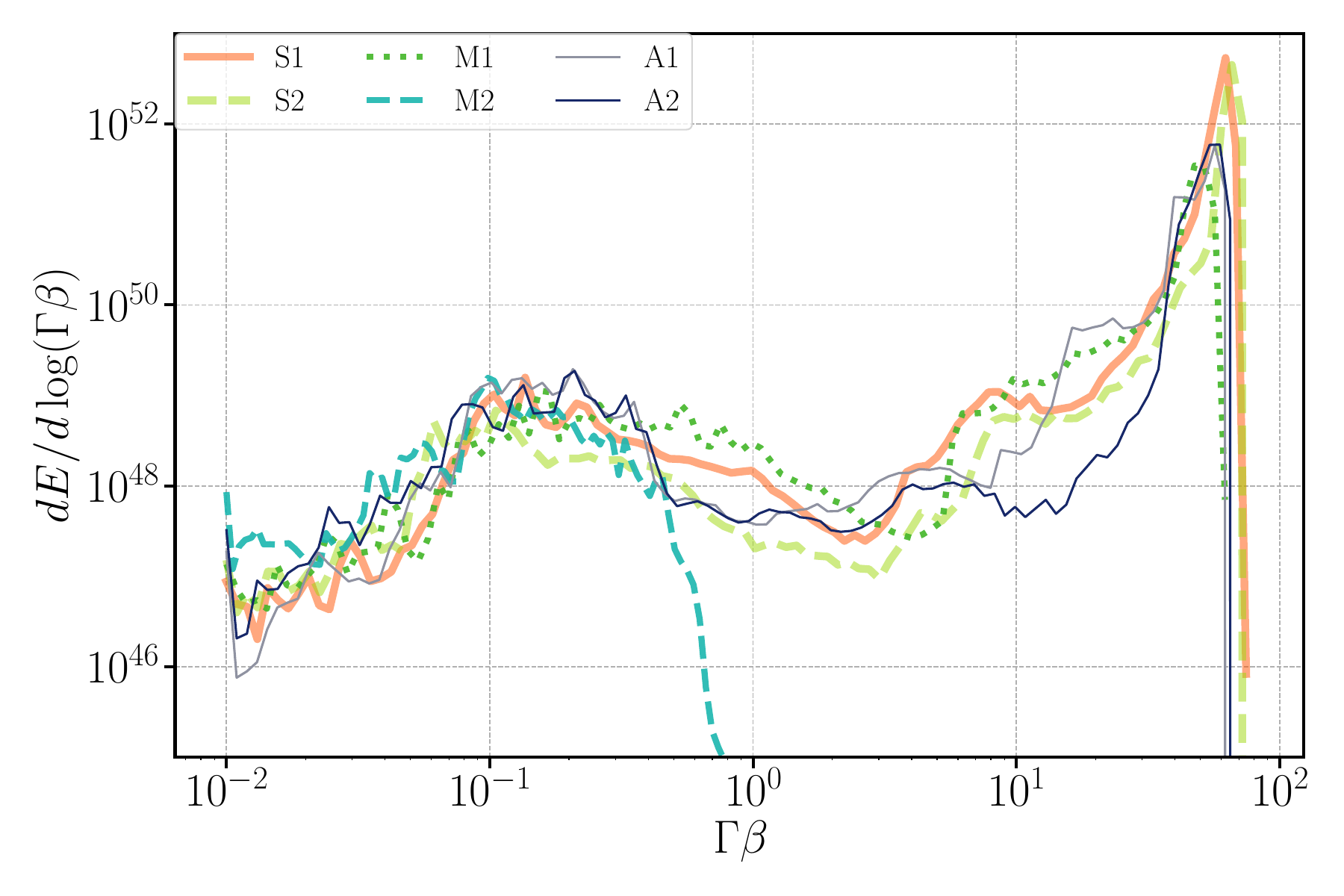}
    \caption{Energy distribution $dE/d\log(\Gamma\beta)$ as a function of four-velocity $\Gamma \beta$ at $t=20$ s. The two components in the distribution, observed in all models except M2, correspond to mildly relativistic expanding material associated with the jet cocoon and to the highly relativistic jet component. The energy and velocity of the expanding material determine the GW signal (see equations~\ref{eqn:new_h+} and~\ref{eqn:new_hx}).
}
    \label{fig:EvsVel}
\end{figure}

\section{Jet dynamics}
\label{appB}

The gravitational wave signal depends on both the velocity and energy of each computational cell (see equations \ref{eqn:new_h+} and  \ref{eqn:new_hx}).

Figure~\ref{fig:shock_beta} shows the position and velocity of the jet head for the six models. The velocity of the jet shock front is determined by the momentum balance between the jet material and the surrounding stellar material, evaluated in the shock system of reference \citep{Matzner2003,Bromberg2011}.
The velocity is 
\begin{eqnarray}
  v_{\rm sh}= \frac{v_j}{1+\tilde{L}^{-1/2}}\;,
\end{eqnarray}
where $\tilde{L}=\rho_j h_j\Gamma_j^2/\rho_\star$, $v_j$ is the jet velocity, $h_j$ is the jet enthalpy and $\rho_\star$ the stellar density. 
Although $v_j$ is initially imposed as constant, it is modified by the high pressure within the jet and by recollimation shocks \citep[e.g.,][]{lopezcamara2013,harrison18}. In addition, since this pressure depends on the jet luminosity variability, both $v_j$ and $\Gamma_\infty$ are expected to acquire variability as well \citep[e.g.,][]{lopezcamara2016,Gottlieb2020a_intermitent}.

Figure~\ref{fig:shock_beta} shows that all models remain sub-relativistic while the jet propagates through the dense progenitor. 
Models S1, S2, A1, A2, and M1 accelerate substantially as the jet approaches the outer stellar layers and after breakout, when the declining ambient density allows a larger fraction of the injected energy to appear in fast material. The acceleration is regulated by the jet luminosity and the progenitor structure, and differs among the models during the first $\sim 3$~s. Once the jet head reaches $v_{\rm sh}\sim 0.4~c$, the rate of acceleration becomes qualitatively similar among the different models. In the case of model M2, the jet reaches the stellar surface but remains substantially slower. 

An exception is the model M2 that accelerates abruptly during the first $\sim 2~$s, reaching a velocity of $\beta_{\rm sh}\sim 0.2~c$, which then remains nearly constant until $t\approx 12$~s. This corresponds to a weakly relativistic jet, in which the central engine powers the jet long enough for it to break through the stellar surface at $t\approx 5$~s, maintaining a mildly relativistic velocity. This is substantially different from a choked jet model, in which the jet is powered from the central engine for a short time, and the breakout occurs on much longer timescales \citep{urrutia22GW}.

Figure~\ref{fig:EvsVel} shows the energy distribution $dE/d\log(\Gamma\beta)$ as a function of the velocity 4-vector $\Gamma\beta$. All models except model M2 develop a high-velocity component associated with the relativistic jet in addition to slower material (peaking at $\Gamma\beta\lesssim 0.1$) associated with the jet cocoon.

\end{document}